\documentstyle[prl,aps,multicol]{revtex}
\input epsfig.sty
\newcommand{\be}{\begin{equation}}
\newcommand{\la}{\label}
\newcommand{\ee}{\end{equation}}
\newcommand{\bea}{\begin{eqnarray}}
\newcommand{\eea}{\end{eqnarray}}
\begin{document}

\title{Energy focusing inside a dynamical cavity}

\author{K.~Colanero and M.~-C.~Chu}
\address{Department of Physics, The Chinese University of Hong Kong, 
Shatin, N.T., Hong Kong.}
\maketitle

\begin{abstract}
We study the exact classical solutions for 
a real scalar field 
inside a cavity with a wall whose motion is self-consistently determined 
by the pressure of the field itself. 
We find that, regardless of the system parameters,
the long-time solution {\it always} becomes nonadiabatic
and the field's energy
concentrates into narrow peaks, which we explain by means of a simple mechanical 
system. We point out implications for the quantized theory.

\end{abstract}

\begin{multicols}{2} 
\narrowtext

The dynamics of confined cavity fields interacting with the cavity 
wall is of great interest for the understanding of a variety of problems such 
as hadron bag models \cite{bag}, sonoluminescence \cite{sono}, 
cavity QED \cite{qed} and black hole radiations \cite{bh}.
Previous works have mostly approached the problem assuming an externally 
imposed wall motion, neglecting the effects of the radiation pressure, or 
used the adiabatic approximation \cite{cole,meystre}.
In this paper we study, without any approximation, the dynamics of a real 
scalar field inside a cavity,
the wall of which moves according to the combined force of a static potential
$V(R)$ and the field pressure. This system bears important 
resemblances to more complicated ones, such as the Dirac and  
electromagnetic fields,
since they can be partially or completely cast in the form of a wave equation.
Moreover the classical solutions should be a good approximation to the 
quantized fields at least in the case of a large number of field quanta.
As initial condition for the field we always consider a normal mode of the
static cavity. This is in fact a common situation in the study of many physical systems.

We find that in general the system evolves nonadiabatically, and the field
energy concentrates into narrow peaks. This phenomenon can be understood 
with the help of a simple classical mechanical system.

In the present work we use natural units and hence the action ${\cal S}$
is dimensionless as are the velocities. This simply means that, 
although we are dealing with a classical system, for convenience the action is
taken in units of $\hbar$.
In one space dimension and with the field only inside the cavity, 
the system is defined by the action 

\be
{\cal S}=\int_0^t dt^{\prime} \left\{ {1\over 2} M \dot{R}^2 - V(R)+
\int_0^R dx 
{1\over 2} \left[ \phi_{t^{\prime}}^2-\phi_x^2 \right] \right\} .\\
\ee

Imposing $\delta{\cal S}=0$ under any variation of the dynamical 
variables that vanishes at 
$t^{\prime}=0$ and $t^{\prime}=t$ we obtain:

\be
M\ddot{R}+{\partial V(R)\over \partial R}
-{1\over 2}\left[ \phi_t^2-\phi_x^2\right]_{x=R} =0 \ \ ,
\la{eqmtn1} 
\ee

\be
\phi_{tt}-\phi_{xx}=0 \hspace{2 cm} 0\leq x<R \ \ ,
\la{eqmtn2}
\ee

\be
\begin{array}{ll}
\phi_x=0	\hspace{2 cm}& {\rm at} \ x=0 \ \ , \\
\phi_x=-\dot{R} \phi_t \hspace{2 cm}&  {\rm at} \ x=R \ \ .
\end{array}
\la{bndcndt1}
\ee

Notice the dependence on $\dot{R}$ of the boundary 
conditions. If $\phi (R)=0$ is imposed, the total energy, which is 
conserved 
for a static cavity, is no longer constant for $\dot{R}\neq 0$. 
Eq.~\ref{eqmtn2} is satisfied by $\phi(x,t)$ with

\be
\phi(x,t)=G(t-x)+G(t+x) \ \ 
\la{phigpgm}
\ee
and the positive sign between the two $G$'s ensures that the first boundary condition of Eqs.~\ref{bndcndt1}
is met.
Substituting Eq.~\ref{phigpgm} in the second of Eqs.~\ref{bndcndt1} we obtain:

\be
G^\prime(t+R(t))={1-\dot{R}(t) \over 1+\dot{R}(t)} 
G^\prime\left(t-R(t)\right) \ \ .
\la{gpgmbnd}
\ee

For prescribed wall motion, $G(z)$ for any $z$ can be found by 
using Eq.~\ref{gpgmbnd} and the null line method~\cite{cole}.  
It is assumed that the cavity is static for $t \leq t_0$ with a length $R(t_0)$.
This is equivalent to saying that there is a static zone 
$z \leq z_0=t_0+R(t_0)$, in which  $G(z)$ is analytically known.
One can find the values of $G(z > z_0)$ 
outside the static zone by first solving the algebraic
equation $z=t_{\rm eqv}+R(t_{\rm eqv})$ for $t_{\rm eqv}$ and then 
finding $z_- \equiv t_{\rm eqv}-R(t_{\rm eqv})$.  
This process, which is
equivalent to constructing a null line connecting the points $z$ and $z_-$,
can be repeated many times until a point $z_s$ 
in the static zone is reached.
The values of
$G(z)$ and $G(z_s)$ are related through Eq.~\ref{gpgmbnd}.
However in the case under study, we do not have, in general, a static zone, 
and we need to verify that knowing the initial conditions of the system 
is enough to implement the above method.

We will show that in order to find $\phi(x,t+dt)$ with $0\leq x\leq 
R(t+dt)$, 
it is necessary and sufficient to 
know $G(z)$ and $G^\prime(z)$ for $t-R(t)\leq z \leq t+R(t)$ and 
$R(t^\prime)$ for $t\leq t^\prime \leq t+dt$.
That is just what is required in order to have a unique solution of the 
system of two second order equations~(\ref{eqmtn1}) and~(\ref{eqmtn2}).

Since $\phi(x,t+dt)=G(t+dt-x)+G(t+dt+x)$, we need to find $G(z)$ and 
$G^\prime(z)$ for 
$t+dt-R(t+dt)\leq z\leq t+dt+R(t+dt)$. Now we have two cases: 
either $z\leq t+R(t)$ or $z > t+R(t)$. 

In the first case it is also true that 
\begin{displaymath}
z\geq t+dt-R(t+dt)\geq t-R(t)
\end{displaymath}
as long as $\dot{R}\leq 1$, 
{\it i.e.} in all physical situations, so that we already 
have the solution. 

In the second case we have to solve the equation $z=t_{\rm eqv}+
R(t_{\rm eqv})$, as explained previously. We have 
\begin{displaymath}
t+R(t)\leq t_{\rm eqv}+R(t_{\rm eqv}) \leq t+dt+R(t+dt) ,
\end{displaymath}
which, with $\dot{R}\geq -1$, implies 
\begin{displaymath}
t\leq t_{\rm eqv} \leq t+dt . 
\end{displaymath}
Having found $t_{\rm eqv}$ we can derive $G^\prime(z)$ from 
Eq.~\ref{gpgmbnd} because,
with $z_{\rm eqv}\equiv t_{\rm eqv}-R(t_{\rm eqv})$,
\begin{displaymath}
t-R(t)\leq z_{\rm eqv}\leq t+R(t) \hspace{1 cm} |\dot{R}|\leq 1 \ \ ,
\end{displaymath}
so that again we have the necessary information to determine the evolution 
of the field. $G(z)$ can then be obtained by the numerical integration 
of $G^\prime(z)$.
Note however that while $\dot{R}=1$ still
admits a solution for the field, $\dot{R}=-1$ doesn't, because the boundary 
condition
requires $G^\prime\left(t+dt-R(t+dt)\right)=G^\prime\left(t+2 dt-R(t)\right) 
=0$, which in general is inconsistent. Evolving backward in time, 
{\it i.e.} with $dt <0$, the opposite
would be true.

Using the procedure above we have studied the case with 
$V(R)={1\over 2} K (R-R_0)^2$, solving, step by step, Eq.~\ref{eqmtn1} 
numerically by a standard finite difference method.

As initial condition for the field we choose the fundamental mode of 
the static cavity with Eqs.~\ref{bndcndt1} as the b.c., 
$R(t_0)=R_0$, and $\dot{R}(t_0)=0$:
\be
\left\{ \begin{array}{l}
\phi=\sin\omega t_0 \cos\omega x \ \ , \\
\phi_t=\omega \cos\omega t_0 \cos\omega x \ \ ,
\end{array}
\hspace{0.6 cm} w\equiv{\pi\over R_0} \ \ .
\right.
\la{incndtf1}
\ee
For convenience we define the dimensionless parameters $\alpha$ and 
$\beta$:
$\alpha \equiv M/\omega$, $\beta \equiv \Omega/\omega = 
\sqrt{K/M}/\omega$,
and we set the amplitude of the initial field to be 1. 
In the case of a wall initially at rest and with a large mass compared to 
the initial energy of the field, we expect the dynamics not to depart 
considerably from the adiabatic one, that is, the wall's motion should be 
well approximated by the solution of Eq.~\ref{eqmtn1}, with the field's 
pressure term replaced by its 
static wall counterpart and the solution of Eq.~\ref{eqmtn2} by
\be
\phi(x,t)=\sin\omega(t)t \cos\omega(t)x \hspace{0.6 cm} 
w(t)\equiv{\pi\over R(t)}\  .
\la{adbtc}
\ee

In order to check the reliability of our numerical implementation of the 
algorithm, we first considered a large mass of the wall 
($\alpha=1000/\pi$, $\beta=1/(10\pi \sqrt{2})$).
We verified that the total energy is very well conserved 
and the motion of the wall is well reproduced by the solution of 
Eq.~\ref{eqmtn1} with the static wall solution for the field pressure.

\begin{figure}
\psfig{file=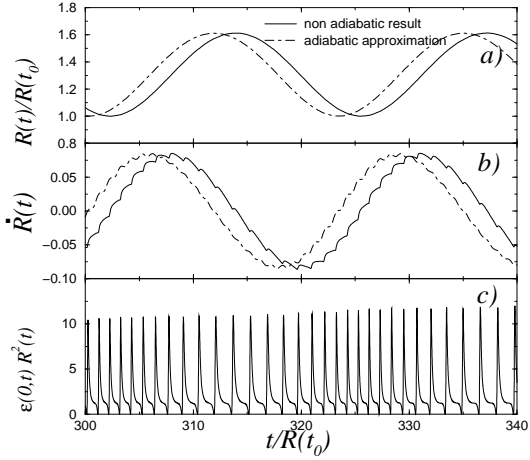,angle=0,width=7 cm}
\caption{a) Wall position, b) wall velocity, c) energy density of the 
field at $x=0$ for $\alpha=100/\pi$, $\beta=1/(\pi \sqrt{20})$. }
\label{fig1}
\end{figure}

\begin{figure}
\psfig{file=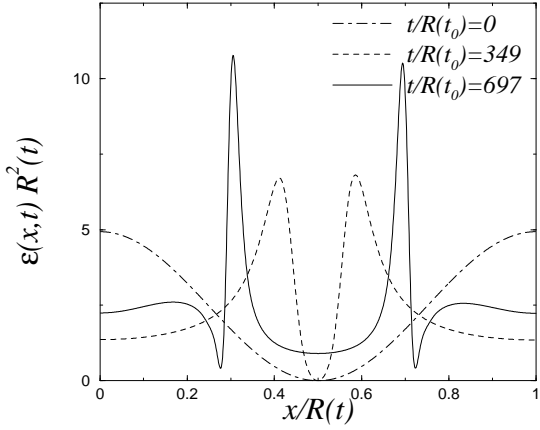,angle=0,width=7 cm}
\caption{Spatial distribution of energy density at 
$t/R(t_o) = 0$ (dot-dashed), 349 (dashed line), and 
697 (solid line) for $\alpha=100/\pi$, $\beta=1/(\pi \sqrt{20})$.}
\label{fig2}
\end{figure}

\begin{figure}
\psfig{file=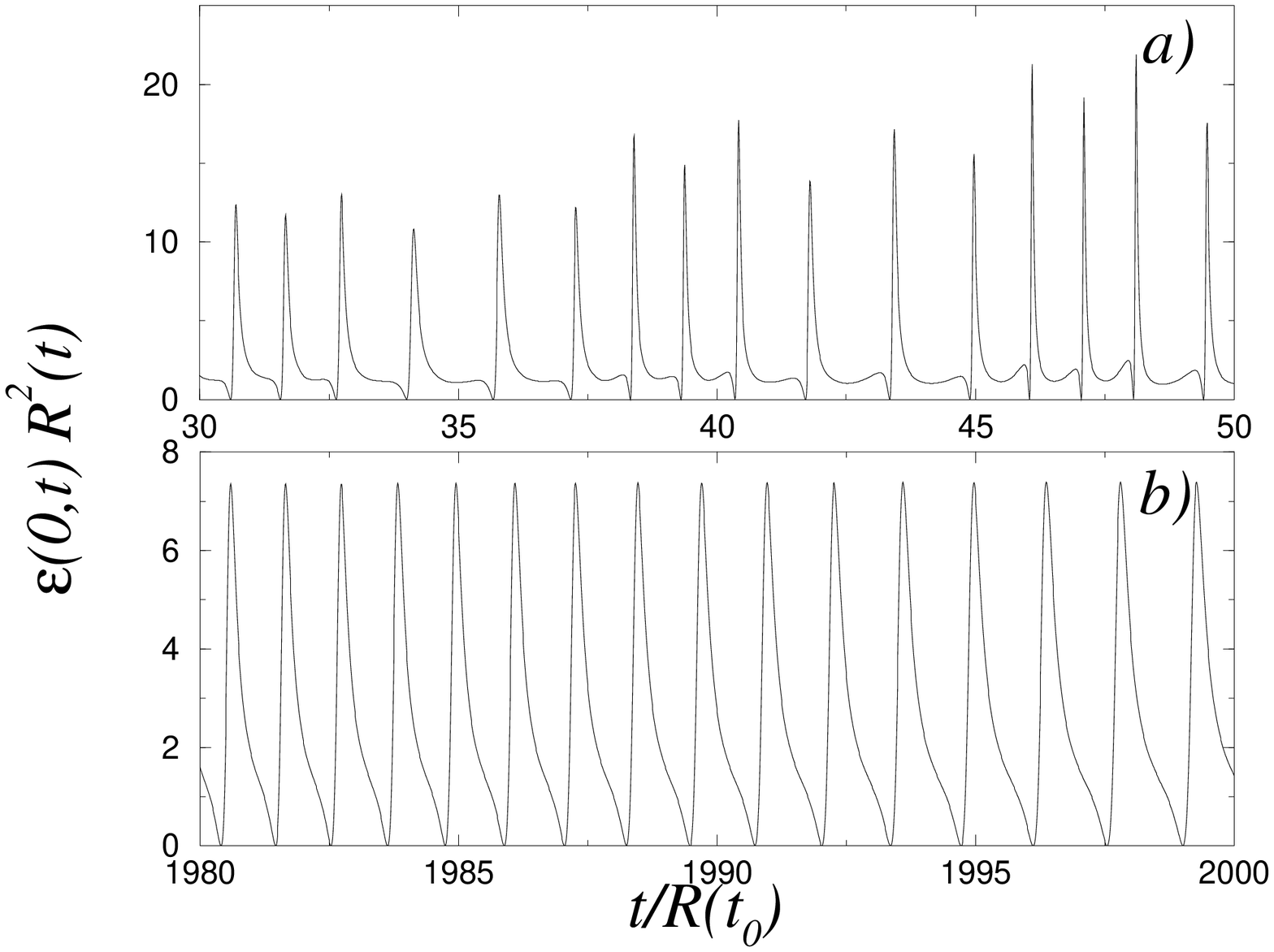,angle=0,width=7 cm}
\caption{Energy density of the field at $x=0$ for: a) 
$\alpha=10/\pi$, $\beta=1/(\pi\sqrt{2})$, b) $\alpha=1000/\pi$, 
$\beta=1/(10\pi\sqrt{2})$. Notice the time intervals.}
\label{fig3}
\end{figure}

We then used a smaller mass keeping $K$ constant, 
{\it i.e.}~$\alpha=100/\pi$ and $\beta=1/(\pi \sqrt{20})$. 
As shown in Fig.~\ref{fig1}, both the wall motion  
and the field energy density 
become nontrivial. An interesting feature is the concentration of the 
energy density, shown in Fig.~\ref{fig1}$c$. This is confirmed by the 
plot of the energy density at two instances $t=349R(t_o)$ and 
$t=697R(t_o)$ in Fig.~\ref{fig2} compared with the static cavity solution. 
The two peaks at $t=697R(t_o)$  move in opposite directions, and their 
widths decrease in time.
This phenomenon is even more evident with $\alpha=10/\pi$ and 
$\beta=1/(\pi \sqrt{2})$ (Fig.~\ref{fig3}a), showing a complex 
distribution of the peak locations and heights.
The total energy of the system is the same in all cases.

Even for the case in Fig.~\ref{fig3}b ($\alpha=1000/\pi$, 
$\beta=1/(10\pi \sqrt{2})$), for which we observed the adiabatic 
evolution lasting for
a long time after $t_0$, we can still, letting the system evolve long enough, 
observe the squeezing of the field energy density 
in spite of the slow motion of the wall. Keeping $K$ constant we found that
the time at which the focusing of the energy starts increases roughly linearly 
with $M$. This suggests that, as 
one takes into account the backreaction of the field on the wall motion, 
the long-time dynamics always becomes nonadiabatic.
We have verified that this remains true also changing the 
boundary conditions so that the field equals zero at the boundaries.

We believe that the origin of this phenomenon lies in the mechanism of 
energy exchange between the wall and the 
field. To explain it we give the following qualitative argument. 
Let's consider the interaction between the wave inside the cavity and the 
wall. At some instance, the peak of the wave will hit the wall, which can be 
moving either outward or inward.  In the former case, 
there will be a transfer of energy from the field to the wall, and the 
speed of the wall will increase slightly. The wavefronts following the 
peak will lose more and more energy to the wall, since the wall moves faster 
with each successive collision.  As a result the spatial width of the energy 
distribution decreases.  When the wall moves inward, the wave gains energy 
from the wall, and the wavefronts following the peak gain less because 
the wall moves slower with each successive collision.  Again the width 
of the waveform decreases. After some time, this effect leads to a 
drastic concentration of energy into narrow peaks.

Our argument depends only on kinematics and should therefore be applicable
not only to waves but many other systems, such as a set 
of particles bouncing back and forth in a dynamical cavity.
For simplicity we consider the dynamics of a set of massless 
non-interacting particles, each having 
momentum and energy $p_i$, $|p_i|$ ($c=1$). 
Inside the cavity they move unperturbed 
at the speed of light. If a particle bounces on the static wall, its 
momentum changes sign.
The movable wall is subjected to a harmonic potential 
$V(R)={1\over 2} K \left( R-R_0 \right)^2$.
The particle momentum $p_i^{\prime\prime}$ and the wall velocity
$v^{\prime\prime}$ after an 
interaction, which is assumed to be instantaneous, are easily derived from
energy and momentum conservation:
\be
\begin{array}{l}
v^{\prime\prime}=\sqrt{(1+v^{\prime})^2+4{p^{\prime}/M}} - 1 \ \ ,\\
p^{\prime\prime}=p^{\prime}+M \left( v^{\prime} - v^{\prime\prime}\right) 
\ ,
\end{array}
\la{parteq1}
\ee
where $v^{\prime}$ and $p^{\prime}$ are the wall velocity and particle
momentum before the collision.
The above equations are derived assuming that the sign of 
$p^{\prime\prime}$ is always
opposite to the sign of $p^{\prime}$, which is true as long as 
the speed of the wall is less than 1 and $2M(1-v^\prime)>p^\prime$ 
($p^\prime > 0$).

\begin{figure}
\psfig{file=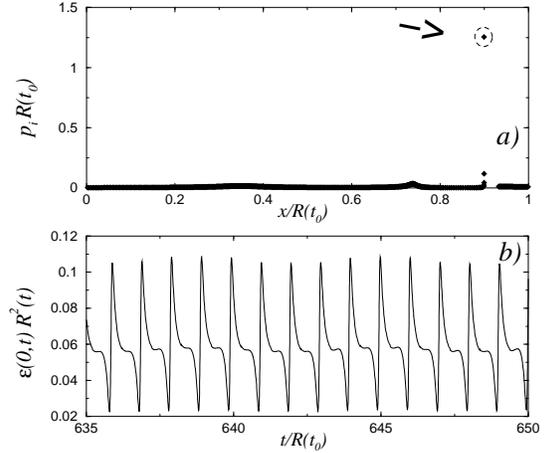,angle=0,width=7 cm}
\caption{Classical particles in a dynamical cavity, with $M=1000/R(t_0)$, 
$\Omega=1/R(t_0)$, and initial momenta $0.01/R(t_0)$. 
a) Particle momenta at $t=3221R(t_0)$.
b) Generalized energy density at $x=0$.}
\label{fig4}
\end{figure}

We consider first a set of $1000$ particles all with the same 
initial momentum $p_i=0.01/R(t_0)$ and
a wall initially at rest with $M=1000/R(t_0)$ and $\Omega=1/R(t_0)$. 
Already after a few interactions with the wall we could 
observe a regular transfer of energy from the last to the first particles
to hit the wall.  In Fig.~\ref{fig4}a we show the 
momenta of the particles after a time $t=3221 R(t_0)$ as a function of 
their position. For clarity only positive momenta are plotted. It is 
remarkable that the first particle to hit the wall has gained more than one 
tenth of the total energy of the system. The above is a very special situation
which however demonstrates the process of energy transfer among particles.

We then extend this simple mechanical model to the case of an 
infinite number of particles
labeled with a continuous index $k$, each having position $q(k)$ and 
momentum $p(k) dk$. In this way we can define an energy density:
\be
{\cal E}(x,t) \equiv \int dk |p(k,t)| \delta(q(k,t)-x) \ \ .
\la{parteq2}
\ee
Not surprisingly ${\cal E}(x,t)$ satisfies the wave equation inside the 
cavity. We numerically simulate such a system choosing $2000$ particles.
Initially, we put two particles at each of the 1000 uniformly separated
sites, and the pairs have opposite momenta 
$p(k)=\pm 10(\pi^2\cos^2\pi q(k)+1)$. 
In Fig.~\ref{fig4}b we plot ${\cal E}(0,t) R^2(t)$, which is evidently 
similar to Fig.~\ref{fig1}, although the details of the evolution 
depend on how the particles or the field interact with the wall.

After a long time we observe the formation of many smaller peaks
in the energy density. Further work is needed to understand 
the problem of the $t\rightarrow \infty$ evolution of the system.

\begin{figure}
\psfig{file=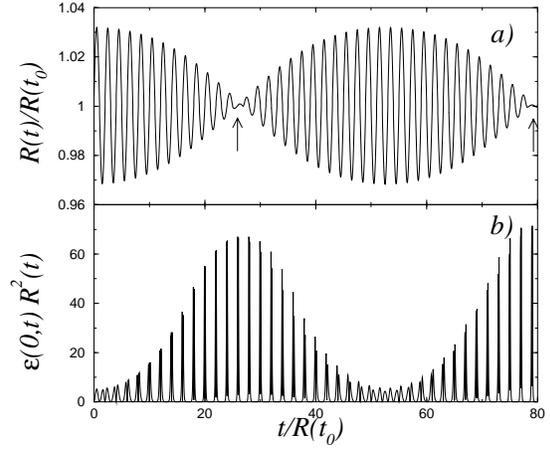,angle=0,width=7 cm}
\caption{a) Wall position and b) energy density in a resonant cavity 
with $\alpha=1000/\pi$ and $\beta=1$.}
\label{fig5}
\end{figure}

For the scalar field an important situation to study is when 
$\Omega=\pi/R_0$, {\it i.e.}, when the wall motion is in resonance
with the field inside the cavity. We have computed the solutions of 
Eqs.~\ref{eqmtn1} and \ref{eqmtn2} for 
various masses of the wall. In Fig.~\ref{fig5} we plot the wall's 
position and the field energy density at $x=0$
vs.~time in the case of $\alpha=1000/\pi$, $R(t_0)=R_0$ 
and $\dot{R}(t_0)=0.1$. In this case we choose
$t_0=R_0/2$ so that $\dot{\phi}=0$ and the initial 
functions Eq.~\ref{incndtf1} satisfy the boundary 
condition Eqs.~\ref{bndcndt1} with $\dot{R}(t_0)\neq 0$. 
Besides the beats in the wall motion, two features
are important. One is the fact that the wall continues to return to 
its initial position after a time $T=R_0$. 
This is different from the case of non-resonant wall parameters 
where the back reaction of the field
changes the frequency of the wall motion. Another remarkable effect, as
a consequence, is
the appearance of narrow peaks typical of a resonantly driven wall 
motion \cite{cole,wai,cklaw}. This indicates the possibility 
of transferring a large amount of energy to the field even with 
an {\it external, non-resonant, driving force}\cite{dodon,jaekel}. 
As long as the frequency of the 
cavity wall is 
$\Omega=\pi/R_0$, it is enough to push the wall at the instances marked by
the arrows in Fig.~\ref{fig5}, and this 
frequency depends on the mass of the wall and can be much smaller 
than $\Omega$; increasing the mass decreases the frequency of energy exchange
between wall and field.
This fact might help to by-pass the experimental difficulty of
achieving a resonant driving force, i.e. at frequency $\Omega$, on a mirror in order to
produce high frequency photons \cite{dodon}.

We have verified that for a small mass, 
$\alpha=10/\pi$, the wall period remains close
to $T=2R_0$ so that the motion is still resonant \cite{wai}.

In Ref.~\cite{wai} it has been shown that the method of null lines 
can also be applied to waves inside an oscillating {\it spherical} 
cavity for any value of the angular momentum.
However, when considering a self-consistent wall motion, the spherical 
symmetry is achieved only in the case of s-waves, for which the radial 
($\phi$) and angular parts can be separated. Defining 
$\psi \equiv r\phi$, so that $\psi$ satisfies the one-dimensional 
wave equation, we can apply the null lines method.
The boundary condition for $\phi$, derived from the action similarly to 
Eqs.~\ref{bndcndt1}, is:
\be
\dot{R}\phi_t(R(t),t)=-\phi_r(R(t),t) \ \ ,
\la{bndcndt3}
\ee
which however for $\psi$ translates to:
\be
\dot{R}\psi_t(R(t),t)={\psi(R(t),t)\over R(t)} - \psi_r(R(t),t) \ \ .
\la{bndcndt4}
\ee
If we want $\phi$ to be finite at $r=0$ then we must require $\psi=0$ at 
$r=0$,
which is satisfied by writing $\psi=G(t-r)-G(t+r)$.
Eq.~\ref{bndcndt4} becomes:
\be
\begin{array}{ll}
&G^\prime\left(t+R(t)\right)- \eta G\left(t+R(t)\right)\\
&= \gamma = -{1-\dot{R}\over 1+\dot{R}}G^\prime\left(t-R(t)\right)-
\eta G\left(t-R(t)\right) \ ,
\la{gpgmbnd2}
\end{array}
\ee
with $\eta \equiv 1/ R\left(1+\dot{R}\right)$. An effective way
to solve Eq.~\ref{gpgmbnd2} {\it numerically} for $G\left(t+R(t)\right)$ is to define
$z=t+R(t)$ and to approximate $\eta$ and $\gamma$ with a constant value between
$z$ and $z-dz$ for a small enough $dz$. Integrating Eq.~\ref{gpgmbnd2}
between $z$ and $z-dz$ we obtain:
\be
G(z)=\left[G(z-dz)+{\gamma\over \eta}\right]e^{\eta dz} - 
{\gamma\over \eta} \ \ ,
\la{3dnl}
\ee
which turns out to be more accurate than standard numerical integration.

\begin{figure}
\psfig{file=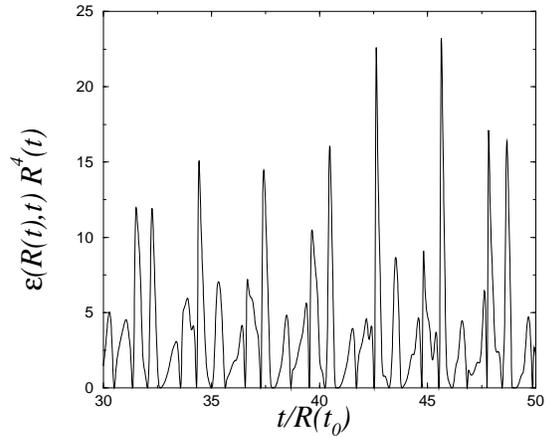,angle=0,width=7 cm}
\caption{Energy density in a spherical cavity at $r=R(t)$ vs.~time for 
$\alpha=5/4.4934$ and $\beta=\sqrt{8}/4.4934$.}
\label{fig6}
\end{figure}

The force of the s-wave field on the wall is
$F_\phi=2\pi R^2(t) \left[ \phi_t^2-\phi_r^2 \right]$.
For $\dot{R}(t_0)=0$ we set as initial conditions for the fields:
\be
\left\{
\begin{array}{l}
\phi(r,t_0)={\cos\omega t_0\over R_0^2}{\sin\omega r\over \omega r} \ \ , 
\\
\phi_t(r,t_0)=-{\omega \sin\omega t_0\over R_0^2}{\sin\omega r\over 
\omega r} \ \ ,
\end{array}
\right.
\la{incond2}
\ee
where $\omega\simeq 4.4934/R(t_0)$ is chosen such that $\phi (r,t_0)$ 
satisfies Eq.~\ref{bndcndt3} with $\dot{R}(t)=0$. 
As in the 1D case, we observe the formation of high energy 
density regions, although in 3D, this process is 
much slower. In Fig.~\ref{fig6} we plot the energy density at 
$r=R(t)$ vs.~time for $\alpha=5/4.4934$ and 
$\beta=\sqrt{8}/4.4934$. These values of the parameters produce a 
completely non-adiabatic evolution. 
For larger $M$ or smaller $K$ we have to evolve the system for a much 
longer time in order to observe the formation of high energy peaks.
However we have verified that imposing $\phi=0$ at $r=R(t)$ the peaks 
appear much earlier and the dynamics is very similar to the one-dimensional 
situation.
With resonant wall parameters,  $\Omega=\pi/R(t_0)$, the 
features observed in 1D remain in 3D. 
With the b.c.~Eq.~\ref{bndcndt3} 
it is also possible to have resonances with $\Omega$ equal to the difference between the 
frequencies of the $n^{\rm th}$ mode 
and the fundamental mode of the cavity. However such an $\Omega$ is close 
to $n\pi/R(t_0)$ if $n$ is large, and such resonances are not easily
distinguishable from the geometric ones \cite{wai}.

In summary we have applied the
null lines method to study the dynamics of a scalar
field inside a cavity whose wall is subjected to a harmonic force and 
the pressure due to the scalar field. 
We have found that the long time evolution of the system
is always non-adiabatic, regardless of the parameters of the system. 
In particular there is an interval of time when the field develops narrow 
packets in energy density 
that bounce back and forth inside the cavity, which can be understood by 
means of a simple mechanical analog 
consisting of a set of massless particles bouncing inside a 
one-dimensional box with a movable wall. Such a system confirms our
hypothesis that the wall motion provides a mechanism of energy transfer 
from low to high energy regions.
We have verified that the focusing of 
energy is a robust phenomenon, being insensitive to the type of potential 
for the wall and the presence of an external driving force. 

For a quantized field previous works \cite{dodon} have shown that in the
case of a prescribed slow wall motion no photon production is achieved. 
Our results strongly suggest that the back-reaction of the field may change 
significantly the evolution of the system. In particular the second 
derivative of the wall position,
which is one of the quantities that determine the number of quanta 
\cite{dodon}, can be much larger than in the adiabatic case, as it can be 
seen from the slope of $\dot{R}$ in Fig.\ref{fig1}. 
If the initial number of fundamental mode quanta is large,
the peaks in energy density in the classical solution can imply 
the production of several high energy quanta.

We have also studied the special situation in which the wall frequency 
is equal to the fundamental
frequency of the static cavity field. Remarkably the frequency of the wall 
motion does not 
change due to the field pressure, and thus narrow peaks typical of a 
resonantly driven wall motion are produced. 
A large amount of energy may be transferred to the field by
providing mechanical energy to the wall when
the amplitude of the oscillation reaches its minimum.
This fact might help to by-pass the experimental difficulty of
achieving a resonant driving force on a mirror in order to
produce high frequency photons \cite{dodon}.

In a further work we would like to address the problem of whether periodical
solutions are admitted for this kind of system and for which values of
parameters.

We would like to thank Dr. C.~K.~Law for his interest in the paper and 
valuable discussions.
This work is partially supported by a Hong Kong Research Grants Council 
grant CUHK 312/96P and a Chinese University Direct Grant
(Project ID: 2060093).

\end{multicols}
\end{document}